\documentclass{PoS}
\usepackage{epstopdf}
\usepackage{latexsym}
\usepackage{amsmath}
\usepackage{amsthm}
\usepackage{fancyhdr}
\usepackage{graphicx}
\usepackage{amssymb,fontenc,times,mathptmx}
\title{On the complex branch points in $\pi N$ scattering amplitude and the multiple poles feature of resonances}

\ShortTitle{Multiple poles feature of $\pi N$ resonances}

\author{\speaker{Shin Nan Yang} \\
        Department of Physics and Center of Theoretical Sciences,\\
        National Taiwan University, Taipei 10617, Taiwan\\
        E-mail: \email{snyang@phys.ntu.edu.tw}}

\abstract{
A simple heuristic argument to
understand the existence of complex branch points in the  $\pi N$ scattering amplitude
is presented.
 It is based on a
hypothesis that the singularity structure of the $\pi N$ scattering
amplitude is a smooth varying function of the pion mass. We then
show that the two-pole structure found to correspond to the Roper
resonance could just a simple direct mathematical consequence of including
additional Riemann surface  in the analysis. Our study
indicates that it is always possible to have multiple poles, either
two or four etc., in different Riemann sheet, to be associated with
a resonance. The poles in all Riemann sheets should be looked for to
determine whether the two-pole feature of the Roper resonance $N^*(1440)$
is a manifestation of the "exact degeneracy" discussed here but
masked by numerical indeterminacy, or an "accidental" one. The determination of the multiplicity
of a pole  could provide some information of
the analytical structure of the numerator of the pole term, as the
numerator of the pole term is related to resonance form factor.}

\FullConference{XV International Conference on Hadron Spectroscopy-Hadron 2013\\
        4-8 November 2013\\
        Nara, Japan }

\begin{document}

\section{Introduction}
Recently, there is a renewal of interest on  the two-pole structure
of the Roper resonance  $N^*(1440)$ unraveled   in the analysis of
$\pi N$ $P_{11}$ partial wave. It was first noted in the SAID
pion-nucleon partial-wave analysis \cite{Arndt85} that, in the
$P_{11}$ channel, in addition to the one reached directly from the
real axis, there is another pole of the scattering amplitude  lying
just behind the complex branch cut corresponding to the opening of
$\pi\Delta$ channel. It led   \cite{Cutkosky90} to
re-examine their previous analysis of \cite{Cutkosky79} and
confirmed the finding of \cite{Arndt85}. They further found that
there are four nearly degenerate poles  corresponding to the
resonance at 1700 MeV,  if poles in other Riemann sheets associated
with $\pi\Delta$ and $\rho N$ branch cuts were searched for.

It is now generally accepted that the two-pole feature of the Roper
resonance is closely connected with the introduction of complex
branch cuts in the analysis. For example, in the DMT meson-exchange
model of $\pi N$ scattering \cite{DMT-PiN}, only one pole was found
to correspond to $N^*(1440)$  as the inelastic $\pi\Delta$ channel
was not explicitly included.

The complex branch point  has been shown to exist using
only the general properties of the S matrix \cite{Ceci11} and
demonstrated to be important for  reliable extraction of resonance
parameters. In this contribution, we present  a simple heuristic
argument to understand the existence of complex branch points for
 $\pi N$ scattering amplitude. It is based on a
hypothesis that the singularity structure of the $\pi N$ scattering
amplitude is a smooth varying function of the pion mass $m_\pi$, or
equivalently the quark masses. We then proceed to show that the
two-pole structure of the Roper resonance  could just be a direct
mathematical consequence when additional Riemann surface is included
in the study and will be a general feature for all resonances when
multi-Riemann sheets are considered. Some of the results presented
here have been reported in \cite{Yang11}.

\section{Existence of complex branch cut
in $\pi N$ scattering} Eden proved  \cite{Eden52} in 1952 that the
elastic scattering matrix element has a singularity at each energy
corresponding to a threshold for a new allowed physical process by
analyzing Feynman amplitudes for any renormalizable field theory.
They all lie on the real axis.

\begin{figure}[h]
\begin{minipage}[t]{70mm}
\begin{center}
\includegraphics[width=1.05\linewidth,angle=0]{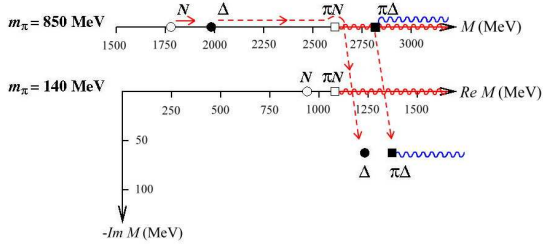}
\end{center}
\caption{Movement of the $\Delta$ pole and the $\pi\Delta$ branch
cut with change of pion mass in the LQCD results if variation w.r.t.
$m_\pi$ would be smooth.} \label{branch-cut}
\end{minipage}
\hspace{\fill}
\begin{minipage}[t]{75mm}
\noindent  horizontal line labeled by $m_\pi=140$ MeV on the left,
in Fig. \ref{branch-cut}. We have purposely aligned the two open
squares corresponding to the $\pi N$ elastic threshold obtained with
$m_\pi=850$ and $m_\pi=140$ MeV, respectively, to show more clearly
how $\Delta$ pole moves as $m_\pi$ is varied. An experimental value
of $\Gamma_\Delta=120$ MeV for the width of the $\Delta$ is assumed
in Fig. \ref{branch-cut} when $m_\pi=140$ MeV.

\hspace{0.3cm}The pole character of the $\Delta$ in the $\pi N$
scattering amplitude would remain, as generally expected, unchanged
after it moves into complex plane, if the singularity structure of
the S matrix would vary smoothly with the pion mass. In the same
\end{minipage}
\end{figure}

Lattice QCD shows that  in the large pion mass region like $m_\pi
\ge 850$ MeV, one always has $ m_\pi + M_N
> M_\Delta > M_N$ such that $\Delta$ is stable.
The $\pi N$ S-matrix would then have   poles at the points
$\sqrt{s}=M_N,$ and $M_\Delta$. Consequently, there would appear two
branch cuts, denoted by the wiggly lines, starting at $m_\pi + M_N$
(open square) and $m_\pi + M_\Delta$ (solid square) on the real
axis, respectively, as shown in the upper horizontal line, labeled
with $m_\pi=850$ MeV on the left, in Fig. \ref{branch-cut}.

The lattice QCD results on the evolution of the masses of the
nucleon (N), $\Delta(1232)$, $\pi N$, and $\pi\Delta$ with pion
mass, as summarized in the Fig. 1 of \cite{Yang11},  indicate that
as $m_\pi$ decreases, $M_N$ and $M_\Delta$  decrease as well,  with
$M_N$ decreasing at a faster pace. On the other hand,  points $M_N$
(open circle) and $M_\Delta$ (solid circle) actually move closer to
$\pi N$ threshold in the process. This is indicated by the arrows on
the solid and dashed lines just above the horizontal line labeled
with $m_\pi=850$ MeV in Fig. \ref{branch-cut}, as $m_\pi$ decreases.
Eventually $M_\Delta$ and $m_\pi + M_N$  would become equal at
around $m_\pi \sim 300$ MeV and $M_\Delta$ would move to the right
of $m_\pi + M_N$, which is the branch point of $\pi N$ elastic cut.
$\Delta$ would then become unstable and begin moving into complex
plane, as shown by the dashed line connecting $\Delta$ on the upper
horizontal line and the $\Delta$ lying in the complex plane below
the lower

\noindent   token, the branch point corresponding to the opening of
$\pi\Delta$ would also move into complex plane and its squared root
character should
  be retained as well. Accordingly, there should exist a branch cut in
the complex plane starting from $m_\pi +M_\Delta$, which is complex
when the value of $m_\pi$ goes down to 140 MeV, as indicated in Fig.
\ref{branch-cut}.
\\

\section{Multiple poles structure of a resonance with several Riemann sheets}

\subsection{Number of Riemann sheets}
  The inclusion of the
$\pi\Delta$ branch cut in the partial wave analysis in $P_{11}$ has
led to the conclusion \cite{Arndt85,Cutkosky90,Suzuki10} that there
are two almost degenerate poles corresponding to the Roper, one on
the unphysical sheet directly reachable from the real axis and the
other located just behind the $\pi\Delta$ cut. We   demonstrate in
the followings that such a two-pole structure could just be a direct
mathematical consequence when there are two Riemann sheets to be
considered.

The $\pi N$ T-matrix is a sum of background $T_B$ and resonances
$T_R$ contributions, i.e., $T_{\pi N}=T_B + T_R$. The background
$T_B$ contains branch points of elastic  and all inelastic cuts.
They appear additively and each of square root nature.  For our
purpose, we will first discuss the question of the number of Riemann
sheets needed to make a function like $T_B$ single-valued in the
presence of several branch points, all of square root nature.

It is  known that a square root function $\sqrt{z}$ requires two
Riemann sheets to make it single-valued. For a general case, we
quote the following mathematical lemma \cite{Wang13}, namely, $2^n$
Riemann sheets are required to make the function
$f(z)=\sum_{i=1}^n\sqrt{z-z_i}$ single-valued. It is easy to see
that the Riemann surface of a function of a more general form
$f(z)=\sum_{i=1}^ng_i(z)\sqrt{z-z_i}$, where $g_i(z)$ are analytic
functions, also contains $2^n$ sheets. Accordingly, the same
conclusion should hold for $T_B$.

\subsection{Multiple poles feature in the presence of branch cuts}
We now turn to the cases where a pole  is present in addition to
branch cuts. Extension to several poles is straightforward. We'll
illustrate our mathematical argument with  explicit examples, all
have a pole at $z_0=r_0e^{i\theta_0}, 0\le\theta_0\le 2\pi$. For simplicity, we'll assume that the
background contribution behaves as $\sum_{i=1}^n\sqrt{z-z_i}$. In
the followings,  $g(z)$ denotes an analytic function and  $a, b,
z_1, z_2$ are constants.

\begin{eqnarray}
  1. \,\, f_1(z)&=&\sqrt{z}+\frac
{g(z)}{z-z_0}. \label{f1} \nonumber\\
 2. \,\, f_2(z)&=&\sqrt{z+z_1}+\frac
{g(z)(\sqrt{z+z_1}\pm\sqrt{z_0+z_1})}{z-z_0}. \label{f2}\nonumber\\
 3. \,\, f_3(z)&=&\sqrt{z+z_1}+\sqrt{z+z_2}+\frac
{g(z)}{z-z_0}.  \label{f3}\nonumber\\
 4. \,\, f_4(z)&=&\sqrt{z+z_1}+\sqrt{z+z_2}+\frac
{g(z)h(z)}{z-z_0}, \,\, h(z)=a\sqrt{z+z_1}+b\sqrt{z_0+z_1}. \label{f4}\nonumber\\
 5. \,\, f_5(z)&=&\sqrt{z+z_1}+\sqrt{z+z_2}+ \frac
{g(z)(\sqrt{z+z_1}\pm\sqrt{z_0+z_1})(\sqrt{z+z_2}\pm\sqrt{z_0+z_2})}{z-z_0}.\nonumber
\label{f5}
\end{eqnarray}

The reason we consider only pole term   of the form of
$N(z)/(z-z_0)$ is that we can always rewrite a function of the form
$\tilde N(z)/D(Z)$, where $D(z)$ has a zero at $z_0$, as
$N(z)/(z-z_0)$ with $N(z)=[\tilde N(z)(z-z_0)/D(z)]/(z-z_0)$. With
some simple mathematical analysis, it is not difficult
\begin{table}[h]
\begin{minipage} {85mm}
\noindent
to reach
  the following conclusions, w.r.t. the numbers of Riemann sheets required
to make the above functions single-valued and the
multiplicity of the pole.\\
1. $f_1(z):$ Two Riemann sheets required. $r_0e^{i\theta_0}$ and $
r_0e^{i(\theta_0+2\pi)}$, located in sheet I and II, respectively,
are  both poles of $f_1(z)$ with equal
residue.\\
2. $f_2(z):$ Two Riemann sheets. Pole
appears only in {\it one} Riemann sheet.\\
3. $f_3(z):$ Four Riemann sheets. Pole   appears in every sheet
with equal residue.\\
4. $f_4(z):$ Four Riemann sheets. Pole   appears in every sheet but
all with {\it different} residue. If   $a = \pm b$, then pole
 appears  only in {\it two} sheets.\\
5. $f_5(z):$ Four Riemann sheets. Pole  appears only in {\it one}
 sheet.
\end{minipage}
\hspace{\fill}
\begin{minipage}[h]{60mm}
\hspace{0.3cm}
\begin{center}
\begin{tabular}{|c||c|c|}
\hline
          Sheet             & Position                        & Residue                        \\
\hline
  I & 1384 - 119$i$ &  19 - 68$i$   \\
  II& 1382 - 138$i$&   -6 - 105$i$    \\
  III& 1513 - 54$i$&   -15 - $i$   \\
  IV& 1514 - 66$i$&   -38 - 5$i$ \\
  & &   \\
  I & 1689 - 54$i$&   -10 + 3$i$  \\
  II & 1676 - 83$i$&  -54 + 21$ i$  \\
  III & 1690 - 61$i$&   -8 + 0$ i$  \\
  IV & 1683 - 83$i$&   -39 + 12$ i$  \\
\hline
\end{tabular}
\caption{The pole positions (MeV) and residues obtained in
\cite{Cutkosky90} by analyzing the VPI data of \cite{Arndt85}.}
\end{center}
\end{minipage}
\end{table}

\noindent

We are hence led to the conclusion that for a complex function with
$n$ square root branch cuts, a pole could appear in all $2^n$
Riemann sheets or only some ($1, 2, 4, 2^{n-1}$) of them, with equal
or different residues. The multiplicity of the pole would reveal some information about
the analytical structure of the numerator $N(z)=\tilde N(z)(z-z_0)/D(z)$
\section{Understanding the multiple-pole structure in $\pi N \,\,P_{11}$ channel}
  Table 1 gives the pole positions (MeV) and residues
obtained in \cite{Cutkosky90} by analyzing the VPI data of
\cite{Arndt85}, where sheet I refers to the sheet most directly
reached from real axis; sheet II is behind the $\pi\Delta$ branch
cut, sheet III is behind the $\rho N$ branch cut, while sheet IV is
behind both  unstable-particle branch cuts. It is seen that pairs of
poles, $(1384 - 119i, \,\,1382 - 138i), \, (1513 - 54i, \,\,1514 -
66i), \,
 (1513 - 54i, \,\,1514 - 66i), \, (1689 - 54i, \,\,1690 - 61i), \, (1676 - 83i, \,\,
 1683 - 83i)$ are all almost degenerate. In fact, the last two pairs are not very
 different from each other as well. The first pair $(1384 - 119i, \,\,1382 -
 138i)$ was first noted by the SAID group \cite{Arndt85} and recently
 received considerable attention from another perspective, e.g., see \cite{Suzuki10}.
 It will be very interesting to see whether those "near degeneracies" are accidental or
 just manifestations of exact degeneracy, as discussed in the above, but masked by numerical
 indeterminacy. In any case, the exact number of degeneracy associated with any resonance
 as discussed here should be determined since it will give us some information about the numerator $N(z)$
 which is related to the form factor of the resonance.

\section{Summary}

In summary, we   present a simple heuristic argument to
understand the existence of complex branch points in the  $\pi N$ scattering amplitude.
 The reasoning is based on a
hypothesis that the singularity structure of the $\pi N$ scattering
amplitude is a smooth varying function of the pion mass. We then
show that the two-pole structure found to correspond to the Roper
resonance could just be a direct mathematical consequence of including
additional Riemann surface  in the study. In fact, our
analysis indicates that it is always likely to find multiple
poles, either two or four etc., to be associated with a resonance.
It is useful to look for pole in all Riemann sheets  to determine
its multiplicity. It would provide us some information of the
analytical structure of the numerator of the pole term, as the
numerator of the pole term is related to resonance form factor.

\section*{Acknowledgment}
I acknowledge gratefully the beneficial discussions  with Profs.
Kuo-Shung Cheng, Chin-Lung Wang, and Wei-Zhe Yang, all
mathematicians.  This work is supported in part by the National
Science Council of ROC (Taiwan) under grant No.
NSC101-2112-M002-025.

\end{document}